\documentclass[aps,12pt]{revtex4}

\usepackage{epsfig}
\usepackage{graphicx}
\usepackage{amsmath,amssymb}
\usepackage{color}

\parskip=\medskipamount

\newcommand{\eq}[1]{(\ref{#1})}
\newcommand{\fig}[1]{Fig. \ref{#1}}

\newcommand{\be}{\begin{equation}}
\newcommand{\ee}{\end{equation}}

\newcommand\disp{\displaystyle}

\newcommand\eps{\varepsilon}

\newcommand{\la}{\left<}
\newcommand{\ra}{\right>}

\begin{document}

\title{On statistical models on super trees}

\author{A.S. Gorsky$^{1,2}$, S.K. Nechaev$^{3,4}$, and A.F. Valov$^5$}

\affiliation{$^1$Institute for Information Transmission Problems of RAS, Moscow, Russian Federation \\ $^2$Moscow Institute of Physics and Technology, Dolgoprudny, Russian Federation \\ $^3$Interdisciplinary Scientific Center Poncelet (CNRS UMI 2615), Moscow, Russian Federation  \\
$^4$P.N. Lebedev Physical Institute of RAS, Moscow, Russian Federation \\
$^5$N.N. Semenov Institute of Chemical Physics of RAS, Moscow, Russian Federation}

\begin{abstract}
We consider a particular example of interplay between statistical models related to CFT on one hand, and to the spectral properties of ODE, known as ODE/IS correspondence, on the other hand. We focus at the representation of wave functions of Schr\"odinger operators in terms of spectral properties of associated transfer matrices on "super trees" (the trees whose vertex degree  changes with the distance from the root point). Such trees with varying branchings encode the structure of the Fock space of the model. We discuss basic spectral properties of "averaged random matrix ensembles" in terms of Hermite polynomials for the transfer matrix of super trees. At small "branching velocities" we have related the problem of paths counting on super trees to the statistics of area-weighted one-dimensional Dyck paths. We also discuss the connection of the spectral statistics of random walks on super trees with the Kardar-Parisi-Zhang scaling.
\end{abstract}

\maketitle

\section{Introduction}

There is an interesting correspondence pioneered in \cite{tateo,blz,suzuki}, between solutions of some statistical models (SM) related to conformal field theories (CFT) and spectral properties of ordinary differential equation (ODE) -- see \cite{rev} for review. This correspondence uncovers the hidden relation between statistical properties of some 2D systems like six-vertex or Potts models with the spectral properties of the quantum mechanical (QM) 1D Schr\"odinger equation in the potential $V(x)$. In the simple cases one has
$$
V(x)=x^{2m} +\frac{l(l+1)}{x^2}
$$
The correspondence between two systems involves establishing dictionary of parameters and the identification of the solution to the Baxter equation at the statistical side to the spectral determinant at the QM side. The Baxter equation for the transfer matrix (TQ) itself was identified with the equation on Stokes multiplayers at QM side. The counterpart of the fusion hierarchy familiar for SM has been found at QM side as well.

The ODE/QM correspondence can be also formulated purely in terms of the integrable structure behind the relevant conformal field theory (CFT) \cite{blz}. The vacuum expectation value (VEV) of the $Q$-operator in CFT at finite temperature can be connected to the spectral determinant at the QM side. In the CFT/QM correspondence, the QM parameters $(m,l)$ are related to the central charge and the dimension of the operator in the corresponding CFT. It was also demonstrated in the CFT framework that the QM problem is related to the (1+1)D Brownian motion in the constant external field with an additional periodic potential \cite{blz2}. The spectral determinants define the VEV of some operators in the stochastic framework.

In this note we reverse the logic traced above, providing the na\"ive example which demonstrates how the wave function of a quantum mechanical problem in a specific non-uniform target space is connected to the spectral determinant at a SM/CFT side and GOE ensemble. The determinant representation of the wave functions in the quantum mechanics has been recently recognized in the matrix model setup \cite{vafa,krefl1,krefl2}, where it was found that the loop equation of the $\beta$-ensemble matrix model with the suitable potential in the Nekrasov-Shatashvili limit yields Schr\"odinger equation and the determinant representation for the wave function.

We show in this paper that the determinant representation of the Hermite polynomials is closely related to the characteristic polynomial for the transfer matrix on a "super tree" whose vertex degree changes linearly with the distance from the tree root. We discuss mapping between the spectral determinant of the transfer matrix and the oscillator wave functions. When the "vertex degree velocity" (the branching increment between neighboring tree levels) is small, we identify the corresponding model with the (1+1)D lattice random walk in the transverse constant magnetic field. Apparently the same problem can be formulated in the symmetric Riemann space with the non-constant radially-dependent curvature, which is the generalization of the space of constant negative curvature (the hyperbolic space).

It is well-known that partition function can be equally derived via the path integral over the trajectories in the coordinate/phase space, or can be represented as a weighted sum over the Hilbert space of energy eigenfunctions. The wave function can be written as the path integral with the fixed extremities and it can be similarly derived via the "path integral" in the Hilbert space.
However the notion of the "path representation over the Hilbert space" has to be precisely specified. That is, we need to define which kind of paths on which kind of a tree in the Hilbert space are relevant for the wave function representation. For the simplest example of the oscillator wave function we argue that the sum over the paths on super trees are relevant for such wave function representation.

To broader extend, our problem seems to be analogous to the problem discussed in the context of the many-body localization \cite{kamenev}. In that work the wave function of the interacting many-body system in the coordinate space is approximated by the sum over the paths having a single degree of freedom on the effective Bethe tree which mimics the Hilbert Fock space of the many-body system. The node of the Bethe tree represents the element in the Fock space, while links between the nodes count non-vanishing matrix elements between the particular states, provided by the interaction term. This picture, presented in \cite{kamenev} though very approximative, allowed however to make important claims about the localization properties of many-body interacting systems \cite{basko}. In our study, the representation of the wave function via the trees in the Hilbert space is exact, but we deal with the oversimplified case of the particle in the external field. Besides, the useful lesson from our study is that the relevant Fock space even for a simple oscillator case, is essentially more complicated than the Bethe tree with the constant branching.

Another question of similar nature deals with the Kardar-Parisi-Zhang (KPZ) scaling, familiar for many growth problems. Usually KPZ scaling emerges in the interacting many-body systems in the physical space. Once again, we can reverse the logic and ask if the KPZ scaling can be recognized in the one-particle propagation problem in the Fock space. It turns out that indeed the answer is positive.

Intensive study of extremal problems of correlated random variables in statistical mechanics has gradually lead mathematicians, and then, physicists, to the understanding that the Gaussian distribution is not as ubiquitous in nature, as it was supposed over the centuries, and shares its omnipresence with another distribution, known as the Tracy-Widom (TW) distribution. The important signature of the TW law is the scaling exponent, $\nu$, of the second moment of the distribution, which is known as the KPZ exponent. For the first time this critical exponent was determined in the seminal paper \cite{kpz} (see \cite{halpin} for review) in the non-equilibrium one-dimensional directed stochastic growth process.

The breakthrough in understanding the ubiquity of KPZ statistics is connected with the works \cite{johansson,spohn} where it was realized that for flat initial conditions the distribution of a rescaled surface height, $\tau^{-1/3}(h(i,\tau)-2\tau)$, in a polynuclear growth converges as $\tau\to\infty$ to the Tracy-Widom (TW) distribution \cite{tw}, providing the statistics of edge states of random matrices belonging to the Gaussian Orthogonal Ensemble (GOE). In the droplet geometry the statistics of growing surface instead corresponds to the edge states of the Gaussian Unitary Ensemble (GUE) \cite{spohn}. Simultaneously, it has been realized that the TW distribution describes the statistics of the ground state energy of an one-dimensional directed polymer in a random Gaussian potential and shortly later the Tracy-Widom distribution was re-derived using the replica formalism typical for disordered systems with the quenched uncorrelated disorder. Below we ask a natural question: could we see some other incarnations of a KPZ statistics besides the extremal events in the nonequlibrium growth? Namely, we perfectly know that the Gaussian exponent, $\nu=\frac{1}{2}$, appears as a critical exponent in the second-order phase transition in the dependence $\xi\sim \tau^{-\nu}$, where $\xi$ is the correlation length and $\tau$ measures the proximity to the critical point (at which $\tau=0$). So, the question is whether there are critical systems which share the dependence like $\xi\sim \tau^{-\nu}$ with $\nu=\frac{1}{3}$?

The paper is structured as follows. In Section II we remind the matrix model representation of the wave functions. In Section III we relate path counting on the super-growing tree with the unit "branching velocity" to the conventional determinant representation for Hermite polynomials. In Section IV the paths counting on the generalized super trees with small branching velocity is compared with the statistics of one-dimensional area-weighted Dyck paths. In the Section V the connection of our model to the "averaged matrix ensemble" in the Edelman-Dumitriu formulation is mentioned. Some open issues are discussed in the Conclusion where we specilate anout possible application of our study to the 1D Anderson localization.

\section{Quantum mechanics from the $\beta$-ensemble}

In this section we remind the determinant representation of the wave functions in quantum mechanics following the line of reasoning formulated in \cite{vafa,krefl1,krefl2}. Consider the integral over the eigenvalues of $N\times N$ matrix $M$
\be
Z=\int \prod_{i}^N d\lambda_i\, (\lambda_i -\lambda_j)^{\beta}\, e^{\frac{\beta}{g_s}W(\lambda_i)}
\ee
where the "weight function" $W(x)$ is called usually the "superpotential". It has been argued in  \cite{krefl1,krefl2} that the matrix element of the operator
\be
\la \det(M-x)^{\beta}\ra = Z^{-1}\int \prod_{i}^N d\lambda_i\, (\lambda_i -\lambda_j)^{\beta} \det(M-x)^{\beta}\,  e^{\frac{\beta}{g_s}W(\lambda_i)}
\ee
in the large--$N$ limit plays the role of the wave function of the Shr\"odinger equation. Specifically, at $N\to \infty$, one sets $g_s\beta N = {\rm const}$, $\beta N = {\rm const}$ and the explicit form of the Schr\"odinger equation becomes
\be
\hbar^2 \frac{d^2\Psi(x)}{dx^2} = \left[\left(\frac{d W(x)}{dx}\right)^2-f(x)\right]\Psi(x)
\ee
where the function $f(x)$ is the polynomial of degree $(d-2)$ if $W(x)$ has degree $d$. The function $f(x)$ is defined as
\be
f(x)=\hbar\big(W'(x) + 2c(x) + d(x)\big)
\ee
with
\be
c(x)= \lim_{\beta \rightarrow 0} \left[\beta N \frac{W'(x)-W(0)}{x}\right], \qquad
d(x)= \hbar \lim_{\beta \rightarrow 0}\left[\beta \hat{D} Z\right]
\ee
where $W'(x) = \frac{dW(x)}{dx}$ and the operator $\hat{D}$ acts on the parameters of the polynomial superpotential $W(x)$ \cite{krefl1}. Effectively, all Virasoro constraints are combined into a single equation. If $\beta$ is finite, the non-stationary Schr\"odinger equation with the same potential emerges.

From the geometric viewpoint one considers the refined topological string in the Calabi-Yau geometry parametrized by the superpotential $W(x)$ in the IIB model. The matrix model microscopically describes the refined topological string, for which the $\beta$-parameter of the matrix model is identified as $\beta=-\frac{\epsilon_1}{\epsilon_2}$ where $\epsilon_1$ and $\epsilon_2$ are the standard equivariant parameters of the $\Omega$--deformation. We are interested in the limit $\beta\rightarrow 0$, which is the Nekrasov-Shatashvili limit of the refined topological string. It was recognized long time ago \cite{aga} that the operator $\det(M-z)$ in the matrix model corresponds to the insertion of the Lagrangian brane in the Calabi-Yau geometry. Hence, from the geometric viewpoint we are dealing with the wave function of the Lagrangian brane in the particular geometry. In the context of the Liouville theory such operator corresponds to the insertion of the FZZT brane.

The immediate question dealing with the Shr\"odinger equation concerns the identification of the particular energy level in the spectrum. To this aim it is useful to consider the Gaussian potential $W(x)= x^2$, see \cite{krefl1}. The spectrum of the corresponding oscillator Hamiltonian can be obtained from the matrix model, it reads
\be
E=\hbar\left(\frac{1}{2} + \lim_{\beta \rightarrow 0}[\beta N]\right)
\ee
where one immediately recognizes the energy level $k = \lim_{\beta \rightarrow 0} [\beta N]$. Therefore, to derive the energy level and corresponding determinant representation for the wave function, one has to start from large--$N$ matrices and proceed with the suitable scaling limits.

In the next Section we remind that for the oscillator case, the $k$-th energy level and the corresponding determinant representation for the wave function can be obtained from the finite  $k\times k$ tridiagonal matrix without appealing to the large--$N$ limit. For the oscillator the $k$-th wave function is proportional to the $k$-th Hermite polynomial which has the representation
\be
H_k(x) \propto \det_{k\times k}(\hat{X} - x)
\ee
where $\hat{X}$ is the operator of coordinate in the Fock space basis. The trilinear recurrent relations follows from the quantization of the system in the action-angle $(I,\phi)$--variables for the oscillator, where $x=\sqrt{I}\, \cos\phi$. Upon the quantization, one also has $\phi= i \frac{d}{dI}$ that is the eigenvalue equation
\be
\hat{X}\Psi_x(n) = x\Psi_x(n)
\ee
, the familiar recurrence for the Hermite polynomials yields immediately.

The $n$-dependence for the elements of the tridiagonal matrix is potential-dependent. To illustrate this point, consider the Sutherland potential
\be
H=\frac{1}{2}p^2 + \frac {l(l-1)}{2\sin^2 x} \qquad \hat{H} \psi_n(x) = \frac{n^2}{2} \psi_n(x)
\ee
In this case, the trilinear recurrent relation reads
\be
\psi_n(x)\, \cos x = \frac{1}{2}\left(\psi_{n+1}(x)\, \sqrt{\frac{1-l(l-1)}{n(n+1)}} + \psi_{n-1}(x)\, \sqrt{\frac{1 -l(l-1)}{n(n-1)}}\right)
\ee
which means that the element of the tridiagonal matrix depends inversely on the row number.

\section{Statistics of paths on super trees}

\subsection{Transfer matrix for the super tree}

Consider the following counting problem: given a regular finite tree, ${\cal T}$, compute the partition function, $Z_N(k)$, of all $N$-step trajectories starting at the tree root $(k=0)$ and ending at some tree level, $k$ ($k=0,...,K-1$). If ${\cal T}$ is the standard Cayley tree (or the Bethe lattice) with the constant branching, $p$, in each vertex at all tree levels, then this counting problem has been discussed infinitely many times in the literature in connection with various physical applications ranging from random walk statistics, polymer topology, localization phenomena, to questions dealing with the RG flows, holography and the black hole structure in the quantum field theory. In all mentioned cases, the uniform $p$--branching Cayley tree, is regarded as a discretization of the target space possessing the hyperbolic geometry -- the Riemann surface of the constant negative curvature.

\begin{figure}[ht]
\epsfig{file=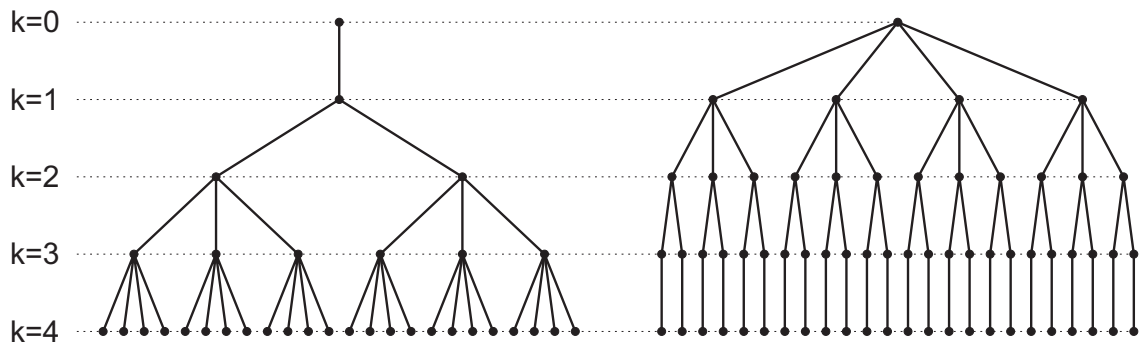,width=15cm}
\caption{Super trees: (a) growing tree ${\cal T}^+$ with $p_0$ branches at the root point and $a=1$, (b) descending tree ${\cal T}^-$ with $p_0=4$ branches and $a=-1$.}
\label{fig:01}
\end{figure}

In this work we deal with the paths statistics on symmetric finite "super trees", ${\cal T}^+$ and ${\cal T}^-$, of $K$ levels, for which the branching (vertex degree) $p$ is not constant, but linearly depends on the current level, $k$ ($k=0,1,2,...,K-1$), i.e.
\be
p_k=\begin{cases} p_0, & \mbox{for $k=0$}  \medskip \\ 2+ak, & \mbox{for $k\ge 1$, $a\ge 0$} \end{cases}
\label{01}
\ee
for "growing trees", ${\cal T}^+$, and
\be
p_k=\begin{cases} p_0, & \mbox{for $k=0$}  \medskip \\ p_0+ak, & \mbox{for $k\ge 1$, $a\le 0$} \end{cases}
\label{01b}
\ee
for "descending trees", ${\cal T}^-$, where "branching velocity", $a$, is some integer-valued constant, and $p_0$ is the branching at the tree root, which is labelled by the index $k=0$. The extension of $a$ to the set of real numbers will be discussed as well. The trees ${\cal T}^{\pm}$ are naturally to identify in the continuum limit with the symmetric Riemann spaces of non-constant negative curvature. The growing tree, ${\cal T}^+$, with $p_0=1$ branches at the root point, and $a=1$, is shown in \fig{fig:01}a, while the descending tree, ${\cal T}^-$, with $p_0=4$ branches at the root point, and $a=1$, is depicted in \fig{fig:01}b.

Before we proceed with a partition function derivation, some important comment dealing with the paths statistics on nonhomogeneous graphs should be made. Since the branching of the tree is not constant, we distinguish between the "path counting" (PC) problem and a more usual "random walk" (RW) statistics. The difference between PC and RW consists in different normalizations of the elementary step: for PC all steps enter in the partition function with the weight one, while for symmetric RW, the step probability depends on the current vertex degree, $p$: the probability to move along each graph bond equals $p^{-1}$. For graphs with a constant $p$ the PC partition function and the RW probability distribution differ only by the global normalization constant, and corresponding averages are indistinguishable. However, for inhomogeneous graphs, like super trees ${\cal T}^{\pm}$, the distinction between PC and RW is crucial: in the path counting problem "entropic" localization of the paths may occur at vertices with large $p$, while it never happens for random walks. The distinction between PC and RW, and the entropic localization phenomenon were first reported for self-similar structures in \cite{17} and later were rediscovered for star graphs in \cite{ternovsky}. More recently this phenomenon was studied in \cite{burda} for regular lattices with defects, where authors introduced a notion of a "maximal entropy random walk" which is essentially identical to the path-counting problem. On a tree with one heavy root the localization phase transition in the path counting problem has been reported in \cite{heavy}.

\begin{figure}[ht]
\epsfig{file=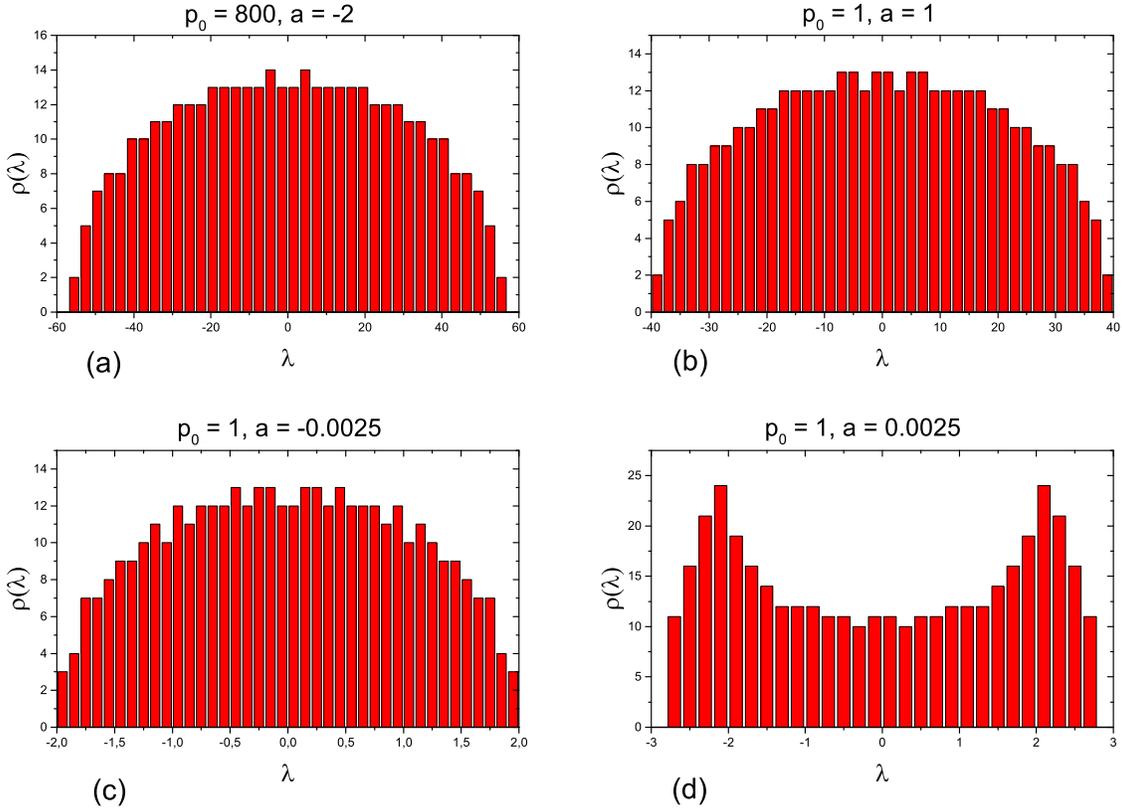,width=15cm}
\caption{Samples of transfer matrices spectral densities for trees of $K=400$ generations and various cases: (a) $p_0=800, a=-2$; (b) $p_0=1, a=1$; (c) $p_0=1, a=-0.0025$; (d) $p_0=1, a=0.0025$.}
\label{fig:02}
\end{figure}

For a growing tree, the partition function, $Z_N(k)$, defined above, satisfies the recursion ($k=0,1,...,K-1$):
\be
\begin{cases}
Z_{N+1}(k)=(p_{k-1}-1)Z_N(k-1)+Z_N(k+1) & \mbox{for $2\le k \le K-1$} \medskip \\
Z_{N+1}(k)=Z_N(k+1), & \mbox{for $k=0$} \medskip \\
Z_{N+1}(k)=p_{k-1} Z_N(k-1)+Z_N(k+1) & \mbox{for $k=1$} \medskip \\
Z_{N+1}(k)=(p_{k-1}-1) Z_N(k-1), & \mbox{for $k=K-1$} \medskip \\
Z_{N=0} = \delta_{k,0}
\end{cases}
\label{02}
\ee
To rewrite \eq{02} in a matrix form, make a shift $k\to k+1$ and construct the $K$-dimensional vector $\mathbf{Z}_N=(Z_N(1),Z_N(2),... Z_N(K))^{\top}$. Then \eq{02} sets the evolution of $\mathbf{Z}_N$ in $N$:
\be
\mathbf{Z}_{N+1}=\hat{T}\mathbf{Z}_N; \qquad \hat{T}= \left(\begin{array}{cccccc}
0 & 1 & 0 & 0 &   \ldots & 0 \medskip \\ p_0 & 0 & 1 & 0 &  &  \medskip \\
0 & p_1-1 & 0 & 1 &  & \vdots \medskip \\ 0 & 0 & p_2-1 & 0 &  & \medskip \\
\vdots &  &  &  & \ddots &  \medskip \\ 0 &  & \dots & & p_{K-2}-1 & 0  \end{array}\right); \qquad \mathbf{Z}_{N=0}=\left(\begin{array}{c} 1 \medskip \\ 0 \medskip \\ 0 \medskip \\ 0 \medskip \\ \vdots \medskip \\ 0\end{array} \right)
\label{03}
\ee
Now we proceed in a standard way and diagonalize the matrix $\hat{T}$. The characteristic polynomials, $P_k(\lambda)=\det(\hat{T}-\lambda \hat{I})$, of the $k\times k$ matrix $\hat{T}$ satisfy the recursion
\be
\begin{cases}
P_k(\lambda)=-\lambda P_{k-1}(\lambda)-(p_{k-2}-1)P_{k-2}(\lambda), & \mbox{for $3\le k \le K$}
\medskip \\
P_1(\lambda)=-\lambda, \medskip \\
P_2(\lambda)=\lambda^2-p_0
\end{cases}
\label{04}
\ee
with $p_k$ given by \eq{01} or \eq{01b}. The spectral density, $\rho(\lambda)$, is constructed as follows. We solve the equation $P_K(\lambda)=0$ for a given $K$, get the set of eigenvalues $\{\lambda_1,...,\lambda_K\}$ and construct the normalized histogram, which counts the degeneracies of each corresponding eigenvalue. The spectral densities for few different values of $p_0$ and $a$ are shown in \fig{fig:02}. Specifically, we have plotted the $\rho(\lambda)$ for transfer matrices of size $K\times K$ for $K=400$ and the following sets of parameters: $p_0=800, a=-2$ for (a), $p_0=1, a=1$ for (b), $p_0=1, a=-0.0025$ for (c), and $p_0=1, a=0.0025$ for (d).

\subsection {Branching velocity $a=1$ and the oscillator potential}

Now we discuss the analytic solution of \eq{04} for a growing tree ${\cal T}^+$ for a special choice $p_0=1$ and $a=1$, and analyze the corresponding asymptotics of $P_K$. The characteristic polynomials, $P_k(\lambda)$ of the transfer matrix $T$ satisfy the recursion
\be
\begin{cases}
P_k(\lambda)=-\lambda P_{k-1}(\lambda)-(k-1)P_{k-2}(\lambda), & \mbox{for $3\le k \le K$} \medskip \\ P_1(\lambda)=\lambda, \medskip \\ P_2(\lambda)=\lambda^2-1
\end{cases}
\label{05}
\ee
which coincide with the recursion for the so-called monic Hermite polynomials, ${\cal H}_k(\lambda)$, also known as the "probabilists' Hermite polynomials":
\be
P_k(\lambda) \equiv {\cal H}_k(\lambda)=(-1)^k e^{\frac{\lambda^2}{2}}{\frac{d^k} {d\lambda^k}}e^{-{\frac{\lambda^2}{2}}}; \qquad{\cal H}_k(\lambda)=2^{-k/2}H_k(\lambda / \sqrt{2)}
\label{05a}
\ee
where $H_k(\lambda)$ are the standard Hermite polynomials. Hence the eigenvalues of the matrix $T$ of size $K\times K$ (see \eq{03}) are the roots of the monic Hermite polynomial, ${\cal H}_k(\lambda)$. In \cite{Kornyik} it has been shown that the normalized roots of the $K^{th}$ monic Hermite polynomial converge weakly at $K\gg 1$ to the Wigner semicircle,
\be
\rho(\lambda)=\frac{1}{2\pi K}\sqrt{4K-\lambda^2}
\label{06}
\ee
The behavior of monic Hermite polynomials, ${\cal H}_k(\lambda)$, at the spectral edge has been analyzed in \cite{Dominici}. For $\lambda\approx 2\sqrt{K}$, the polynomials ${\cal H}_k(\lambda)$ share the following asymptotics
\be
{\cal H}_K(\lambda)\approx \sqrt{2\pi}\, 2^{-K/2} \exp\left(\frac{K\ln(2K)}{2}-\frac{3K}{2}+\lambda\sqrt{K}\right)
K^{1/6}\mathrm{Ai}\left(\frac{\lambda-2\sqrt{K}}{K^{-1/6}}\right)
\label{07}
\ee
where $\disp {\rm Ai}(z)=\frac{1}{\pi} \int_{0}^{\infty} \cos(\xi^3/3+\xi z)\, d\xi$ is the Airy function. Let $a_1>a_2>a_3>...$ be zeros of the Airy function ($a_i<0$ for all $i$). At $K\gg 1$ the maximal eigenvalue, $\lambda_{max}$, of the transfer matrix \eq{03} has the following leading behavior
\be
\lambda_{max}=2\sqrt{K}+a_1\, K^{-1/6}
\label{08}
\ee
where $a_1\approx -2.3381$. At $N\gg 1$ and $K\gg 1$ the averaged partition function, $Z_N=\sum_{k=0}^K Z_N(k)$, can be estimated as follows:
\be
Z_N(K) \approx (\lambda_{max})^N=\left(2\sqrt{K}+a_1K^{-\frac{1}{6}}\right)^N \approx (4K)^{N/2}\, e^{(a_1/2)NK^{-2/3}}
\label{09}
\ee
where we have used the asymptotic expression \eq{08} for the $\lambda_{max}$ for the maximal eigenvalue of the transfer matrix $\lambda_{max}$.

Let us emphasize that the mean distance,
$$
\la k(N) \ra=\frac{\sum\limits_{k=0}^{K} kZ_N(k)}{\sum\limits_{k=0}^{K}Z_N(k)}
$$
between ends of ensemble of open $N$-step paths on a tree ${\cal T}^+$ at $N\gg 1$ coincides with the path length, $N$, i.e.
\be
\lim_{N\to \infty}\frac{\la k(N)\ra}{N} = 1
\ee
The combinatorial entropy, $S_N$, of the ensemble of $N$-step paths on the tree ${\cal T}^+$ has the following asymptotics at large $N$:
\be
S_N = \ln Z_N(K = N) \approx \frac{N}{2} \ln(4N)  + \frac{a_1}{2} N^{1/3}
\label{09a}
\ee

\begin{figure}[ht]
\epsfig{file=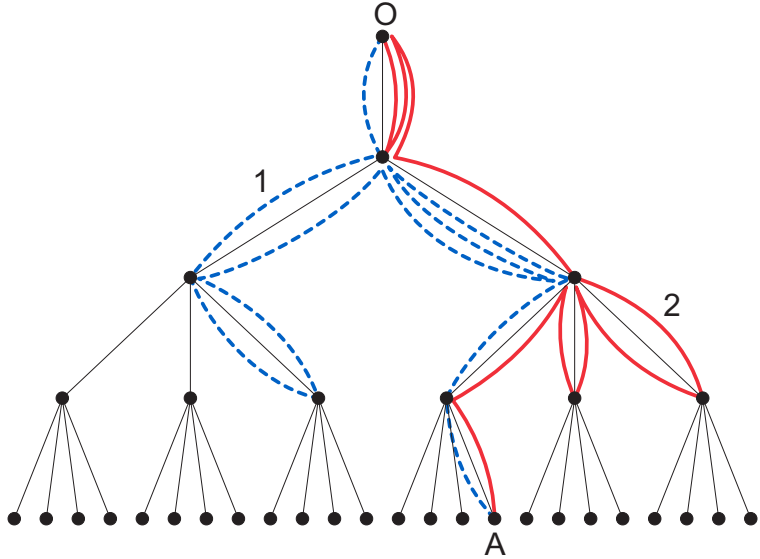, width=10cm}
\caption{Watermelon configuration formed  by two independent trajectories of $N=10$ steps each (solid and dashed). Both trajectories start from the root $0$ and join at the point $A$ on the level $K$ of a growing tree.}
\label{fig:03}
\end{figure}

Now we can estimate the entropy, $S^{(W)}$ of the "watermelon" configuration consisting of two trajectories, 1 and 2, of length $N$ each, both starting at the root point $0$ (as shown in \fig{fig:03}) and meeting each other at the point $A$ located at the distance $K\approx  N$ along a tree. The conditional partition function $Z^{(W)}_N$ of the watermelon configuration can be written as
\be
Z_N^{(W)}= \frac{Z_N\times Z_N}{K!}\bigg|_{K\approx N\gg 1} \approx \frac{(4N)^{N}\, e^{a_1 N^{1/3}}}{\sqrt{2\pi N} e^{-N+N\ln N}}
\label{09b}
\ee
where the denominator in \eq{09b} is the number of vertices at the level $K$ of a growing tree ${\cal T}^+$ and, correspondingly, $1/K!$ is the probability for two terminal points of  trajectories 1 and 2 to meet each other in \emph{one point} at the level $K$, i.e. to form\emph{} a watermelon. The entropy $S_N^{(W)}= \ln Z_N^{(W)}$ reads
\be
S_N^{(W)} \approx c N + a_1 N^{1/3}
\label{09c}
\ee
where $c=2\ln 2+1\approx 2.3863$ and the terms $N\ln N$ in nominator and in denominator of \eq{09b} cancel each other.

It is worth noting that the asymptotic expression \eq{09c} with the finite size corrections controlled by the scaling exponent $\nu=\frac{1}{3}$ appears in our consideration as a simple one-particle paths counting problem on a super tree. Recall that typically the behavior \eq{09c} emerges in \emph{many-body} systems sharing the KPZ statistics, as it has been mentioned in the Introduction. In the model discussed here, we can not claim to receive the full Tracy-Widom distribution, however the KPZ scaling still is accessible. As we show below, our model to some extend can be regarded as a mean-field-like description of a random matrix spectral statistics in the Dumitriu-Edelman setting \cite{edelman}.

\subsection{Paths counting generating functions on supertrees ${\cal T}^+$ and ${\cal T}^-$}

To understand better the connection of the spectrum of polynomials ${\cal H}_k(\lambda)$ with the path counting on super trees, consider the generating function
\be
{\cal Z}(s,k) = \sum_{N=0}^{\infty} s^N Z_N(k)
\label{10}
\ee
Since ${\cal Z}(s,k)\equiv 0$ for $k<0$ and $k>K-1$, rewrite \eq{02} for $a=1$ and $p_0=1$ as follows:
\be
\begin{cases}
s^{-1}{\cal Z}(s,k)=k{\cal Z}(s,k-1)+{\cal Z}(s,k+1) & \mbox{for $1\le k \le K-1$} \medskip \\
s^{-1}{\cal Z}(s,1)={\cal Z}(s,0)+{\cal Z}(s,2) \medskip \\
s^{-1}{\cal Z}(s,0)={\cal Z}(s,1)
\end{cases}
\label{12}
\ee
The recursion is written for a growing tree ${\cal T}^+$ only, the derivation of recursion for a descending tree ${\cal T}^-$ is straightforward.

The function ${\cal Z}(s,k=0)$ is the generating function of all trajectories returning to the root point on a growing finite tree ${\cal T}^+$ of $K$ generations, where each step carries the fugacity $s$. To shorten notations, define ${\cal Z}(s,k=0)={\cal Z}_K^+(s)$. The function ${\cal Z}_K^+(s)$ being the solution of the system of linear equations, can be written as the quotient
\be
{\cal Z}^{+}_K(s)=\frac{\det \hat{B}^+_K(s)}{\det \hat{A}^+_K(s)}
\label{18}
\ee
where
\be
\hat{A}^+_K =\left(\begin{array}{cccccc}
\frac{1}{s} & -1 & 0 & 0 & \ldots & 0 \medskip \\ -1 & \frac{1}{s} & -1 & 0 & \ldots & 0 \medskip \\ 0 & -2 & \frac{1}{s} & -1 & \medskip \\ \vdots & \vdots & \ddots & \ddots & \ddots & \medskip \\
0 & 0 &  & 2-K & \frac{1}{s} & -1 \medskip \\ 0 & 0 &  & 0 & 1-K & \frac{1}{s} \end{array}\right);
\quad \hat{B}^+_K =\left(\begin{array}{cccccc}
\frac{1}{s} & -1 & 0 & 0 & \ldots & 0 \medskip \\ 0 & \frac{1}{s} & -1 & 0 & \ldots & 0 \medskip \\ 0 & -2 & \frac{1}{s} & -1 & \medskip \\ \vdots & \vdots & \ddots & \ddots & \ddots & \medskip \\
0 & 0 &  & 2-K & \frac{1}{s} & -1 \medskip \\ 0 & 0 &  & 0 & 1-K & \frac{1}{s} \end{array}\right);
\label{19}
\ee
One can straightforwardly check that the function ${\cal Z}^{+}_K(s)$ can be expressed in terms of the monic Hermite polynomials and the polynomial $R_k(s)$ which satisfy the recursion $R_{k+1}(s)=R_k(s)-(k+1)s^2R_{k-1}(s)$, where $R_0(s)=R_1(s)=1$:
\be
{\cal Z}^{+}_K(s) = \frac{R_{K-1}(s)}{s^K {\cal H}_K(s^{-1})}
\label{19a}
\ee
Thus, we get
\be
{\cal Z}^{+}(s,k) = \frac{{\cal H}_k(s^{-1})R_{K-1}(s)}{s^K {\cal H}_K(s^{-1})}
\label{20}
\ee
The function ${\cal Z}^{+}(s,K-1)$ is generating function of trajectories starting from the root point and terminating at the end of a growing finite tree ${\cal T}^+$ of $K$ generations
\be
{\cal Z}^{+}(s,K-1)= \frac{R_{K-1}(s^2)}{Ks^K}\frac{d}{d s^{-1}} \ln {\cal H}_K(s^{-1})
\label{23a}
\ee
At $K\gg 1$ and $s^{-1}\approx 2\sqrt{K}$ one gets
\be
{\cal Z}^{+}(s,K-1)\approx \frac{R_{K-1}(s)}{Ks^{K}}\left(\frac{d}{d s^{-1}} \ln \mathrm{Ai}\left(\frac{s^{-1}-2\sqrt{K}}{K^{-1/6}}\right)+\sqrt{K}\right)
\label{23c}
\ee

For a descending tree ${\cal T}^-$ define $p_x=P-x$, where $P$ is the maximal vertex degree of the tree at the root point, and $P$ coincides with the number of tree generations, i.e. $P=K$. The recursions for ${\cal Z}^{-}(s,k)$ on ${\cal T}^-$ read:
\be
\begin{cases}
s^{-1}{\cal Z}^{-}(s,k)=(P-k){\cal Z}^{-}(s,k-1)+{\cal Z}^{-}(s,k+1) & \mbox{for $1\le k \le P-1$} \medskip \\ s^{-1}{\cal Z}^{-}(s,P-2)={\cal Z}^{-}(s,P-1)+{\cal Z}^{-}(s,P-3) \medskip \\
s^{-1}{\cal Z}^{-}(s,P-1)={\cal Z}^{-}(s,P-2)
\end{cases}
\label{24}
\ee
We are interested in computing the canonical partition function ${\cal Z}^{-}(s,k=0)\equiv {\cal Z}^{-}_P(s)$, which enumerates the trajectories returning to the root point on a descending tree, and each step is weighted with the fugacity $s^{-1}$. The function ${\cal Z}^{-}_P(s)$ can be written as (compare to \eq{18})
\be
{\cal Z}^{-}_P(s)=\frac{\det B^{-}_P(s)}{\det A^{-}_P(s)}
\label{27}
\ee
where
\be
\hat{A}^{-}_P =\left(\begin{array}{cccccc}
\frac{1}{s} & -1 & 0 & 0 & \ldots & 0 \medskip \\ 1-P & \frac{1}{s} & -1 & 0 & \ldots & 0 \medskip \\ 0 & 2-P & \frac{1}{s} & -1 & \medskip \\ \vdots & \vdots & \ddots & \ddots & \ddots & \medskip  \\ 0 & 0 &  & -2 & \frac{1}{s} & -1 \medskip \\ 0 & 0 &  & 0 & -1 & \frac{1}{s} \end{array}\right);
\qquad \hat{B}^-_P =\left(\begin{array}{cccccc}
\frac{1}{s} & -1 & 0 & 0 & \ldots & 0 \medskip \\ 0 & \frac{1}{s} & -1 & 0 & \ldots & 0 \medskip \\
0 & 2-P & \frac{1}{s} & -1 & \medskip \\ \vdots & \vdots & \ddots & \ddots & \ddots & \medskip \\
0 & 0 &  & -2 & \frac{1}{s} & -1 \medskip \\ 0 & 0 & & 0 & -1 & \frac{1}{s} \end{array}\right)
\label{28}
\ee
Substituting \eq{28} into \eq{27}, and evaluating the determinants, we get
\be
{\cal Z}^{-}_P(s)=\frac{1}{\disp 1-\frac{(P-1)s}{\disp 1-\frac{(P-2)s}{\disp 1-\frac{(P-3)s}{1-\ldots}}}}= \frac{{\cal H}_{P-1}(s^{-1})}{s {\cal H}_P(s^{-1})}= \frac{1}{Ps}\frac{d}{d s^{-1}}\ln {\cal H}_P(s^{-1})
\label{29}
\ee
At $P\gg 1$ and $s^{-1}\approx 2\sqrt{P}$ the generating function of trajectories returning to the root point, ${\cal Z}^{-}_P(s)$ can be estimated as follows:
\be
{\cal Z}^{-}_P(s)\approx \frac{1}{Ps}\left(\frac{d}{d s^{-1}} \ln \mathrm{Ai}\left(\frac{s^{-1}-2\sqrt{P}}{P^{-1/6}}\right)+\sqrt{P}\right)
\label{33}
\ee
Introducing the new variable
\be
z=\frac{s^{-1}-2\sqrt{P}}{P^{-1/6}}
\label{33a}
\ee
we can rewrite \eq{33} near the spectral boundary, i.e at $s^{-1}\approx 2\sqrt{P}$ as follows
\be
{\cal Z}^{-}_P(s)\approx 2+\frac{2}{P^{1/3}}\frac{d}{dz}\ln \mathrm{Ai}(z) \qquad (z\to 0)
\label{33b}
\ee

\section{Small "branching velocity" and magnetic directed paths}

\subsection{Area-weighted Dyck paths}

In the previous section we have discussed the tree with $a=1$ branching velocity and demonstrated that the dual statistical model corresponds to the oscillator potential and Hermite polynomials. Here we consider the limit of small branching velocity $|a|<<1$. We argue that the random walk on such trees is closely related with the area-weighted Dyck paths where the effective magnetic field on the lattice is related to the branching velocity via a kind of a $T$-duality transform. As a by-product we provide the new interpretation of $q$-Catalan numbers via the super trees.

The problem of a paths counting on super trees is tightly connected with a well-known problem of counting one-dimensional Dyck paths with fixed length and area below the trajectory. Consider a $N\times N$ square lattice and enumerate all $N$-step trajectories (Dyck paths) starting at $(0,0)$, ending at $(N,N)$ and staying above the diagonal of the square (the path can touch the diagonal, but cannot cross it). Let $A$ be the area between the path and the diagonal of the square, counted in full plaquettes. For convenience, turn the lattice by $\pi/4$, as shown in \fig{fig:03}, and consider the partition function of all directed $N$-step paths on a half-line, $k\ge 0$, with the fixed area, $A$, being the sum of all full plaquettes, highlighted in  \fig{fig:03}. Our key object is the area-weighted canonical partition function, $W_N(q)$, defined as follows
\be
W_N(q) = \sum_{\rm Dyck\; paths} q^A
\label{a01}
\ee
where the summation runs over the ensemble of $N$-step Dyck paths enclosing the area $A$, and $q$
is the fugacity of $A$. Writing $q=e^H$, we identify $H$ with a "magnetic field" conjugated to the area $A$.

\begin{figure}[ht]
\epsfig{file=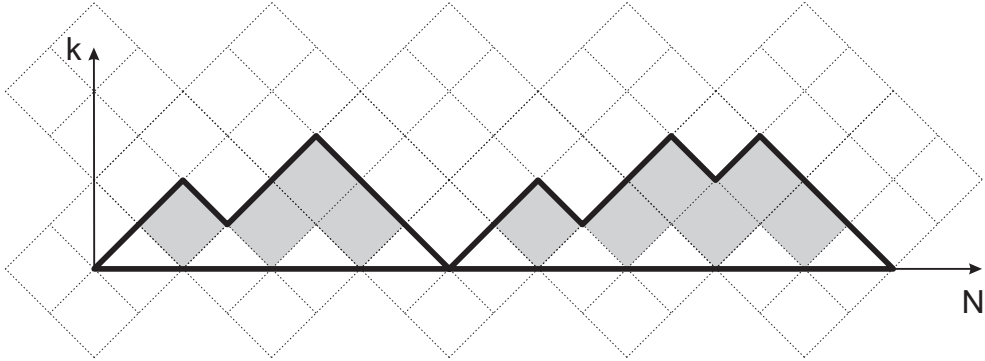,width=13cm}
\caption{The $N$-step Dyck path on a halfline $k\ge 0$ with fixed area below the path measured in full plaquettes.}
\label{fig:04}
\end{figure}

Let us introduce the partition function $W_N(k,q)$ where $k$ is the height of the path at step $N$. The function $W_N(k,q)$ satisfies the recursion
\be
\begin{cases}
W_{N+1}(k,q) = q^{k-1} W_N(k-1,q) + W_N(k+1,q) & \mbox{for $1\le k\le K-1$} \medskip  \\
W_{N=0}(k,q)=\delta_{k,0}
\label{a02}
\end{cases}
\ee
where we have supposed that $W_N(k,q)=0$ for $k\le 0$ and $k\ge K$. The solution of \eq{a02} in a matrix form for a $K$-dimensional vector $\mathbf{W}_N(q)=(W_N(1,q),...,W_N(X,q))^{\intercal}$ (with a shift $k\to k+1$), reads:
\be
\mathbf{W}_{N+1}(q) = \hat{U}(q) \mathbf{W}_N; \qquad \hat{U}(q)=\left(\begin{array}{cccccc}
0 & 1 & 0 & 0 & \dots \\
1 & 0 & 1 & 0 & \\
0 & q & 0 & 1 &  \\
0 & 0 & q^2 & 0 &  \\
\vdots & &  &  & \ddots \end{array}\right); \qquad \mathbf{W}_0=\left(\begin{array}{c} 1 \\ 0 \\
0 \\ 0 \\ \vdots \end{array}\right)
\label{a03}
\ee
Let us note that the value $W_N(1,q)$ defines the partition function of the "Brownian excursion", since at the very last step the trajectory returns to the starting point. Evaluating powers of the matrix $U(q)$, we can straightforwardly check that the values of $W_N(1,q)$ are given by the Carlitz-Riordan $q$-Catalan numbers \cite{carlitz}:
\be
W_N(1,q) = \begin{cases} C_{N/2}(q) & \mbox{for $N=2m$, where $m=1,2,3...$} \medskip \\ 0 & \mbox{for $N=2m+1$, where $m=0,1,2,...$}
\end{cases}
\label{a04}
\ee
Recall that the numbers $C_N(q)$ satisfy the recursion
\be
C_N(q) = \sum_{k=0}^{N-1}q^k C_k(q) C_{N-k-1}(q)
\label{a05}
\ee
which is the $q$-extension of the standard recursion for Catalan numbers. The generating function
$\disp F(s,q)=\sum_{N=0}^{\infty}s^N C_N(q)$ obeys the functional relation
\be
F(s,q) = 1 + s F(s,q) F(sq,q)
\label{a06}
\ee
It is known that the solution of \eq{a06} can be written as a continuous fraction expansion,
\be
F(s,q) =\frac{1}{\disp 1-\frac{s}{\disp 1-\frac{s q}{\disp 1- \frac{s q^2}{1-...}}}}
=\frac{A_q(s)}{A_q(s/q)}
\label{a07}
\ee
where $A_q(s)$ is the $q$-Airy function,
\be
A_q(s)=\sum_{n=0}^{\infty}\frac{q^{n^2}(-s)^n}{(q;q)_n}; \quad (t;q)_n=\prod_{k=0}^{n-1} (1-t q^k)
\label{a08}
\ee
In the works \cite{prel0,rich1,rich2} it has been shown that in the double scaling limit $q\to 1^-$ and $s\to \frac{1}{4}^{-}$ the function $F(s,q)$ has the following asymptotic form (compare to \eq{33b})
\be
F(z,q) \sim {\cal F}_{\rm reg}+(1-q)^{1/3} \frac{d}{dz}\ln {\rm Ai}(4z); \quad
z=\frac{\frac{1}{4}-s}{(1-q)^{2/3}},
\label{eq:asymp1}
\ee
where $F_{\rm reg}$ is the regular part at $\big(q\to 1^-,\, s\to \frac{1}{4}^{-}\big)$. The function ${\cal F}(s,1)$ is the generating function for the non-deformed Catalan numbers:
\be
F(s,q=1)=\frac{1-\sqrt{1-4s}}{2s}
\label{eq:asymp2}
\ee
The generating function $F(s,1)$ is defined for $0<s<\frac{1}{4}$, and at the point $s=\frac{1}{4}$ the first derivative of $F(s,1)$ experiences a singularity which is interpreted as the critical behavior. The limit $q\to 1^-$, $s\to\frac{1}{4}^-$ can be read also from the asymptotic expression for $F(s,q)$. To define the double scaling behavior and derive the Airy-type asymptotic, the simultaneous scaling in $s$ and $q$ is required.

To make connection of area-weighted Brownian excursions to the path counting problem on descending tree, consider the expansion of \eq{a02} at $q\to 1$. Namely we set $q=1-\eps$, where $|\eps|\ll 1$ and expand \eq{a02} up to the first term in $\eps$. We arrive at the following system of equations
\be
\begin{cases}
W_{N+1}(k,q) = \big(1-\eps(k-1)\big) W_N(k-1,q) + W_N(k+1,q) & \mbox{for $1\le k\le K-1$} \medskip \\
W_{N=0}(k,q)=\delta_{k,0}
\end{cases}
\label{a09}
\ee
Comparing equations \eq{02} and \eq{a09} we can note that they are equivalent upon the identification $a= -\eps$ (where $0<\eps\ll 1$) and $p_0=1$. Hence, \eq{eq:asymp1} provides the explicit expression for the path counting on a weakly descending tree with a small non-integer branching velocity, $a$, by expanding the solution to \eq{a09} at $q=1-\eps$ up to the first leading term in $\eps$ and an identification $\eps$ with $a$.

\subsection{"T-duality" and super trees}

Let us make some comments concerning the interplay between the super trees and Dyck paths. The first question concerns the origin of the identification of the branching velocity $a$ for paths on super trees with the fugacity $q$ for (1+1)D magnetic Dyck paths. The parameter $\ln q$ can be naturally associated with the constant magnetic field, $H$, transverse to the lattice, on which the random walk develops. It is also useful to consider this picture as the Euclidean version of the (1+1)D Minkowski space-time with the constant electric field $E$ acting along the space coordinate. Upon the Wick rotation the constant electric field in (1+1)D Minkowski space-time gets transformed into the magnetic field $H$. Such rotation for instance, is used to describe the bounce solution to the Euclidean equations of motion responsible for the creation of the pair in the external electric field.

We have argued above that magnetic field in the Brownian motion picture gets mapped onto the branching velocity for the super tree. The explanation of this mapping is qualitatively provided by a kind of a $T$--duality transform which can be considered as a generalization of a Fourier transform. Let us emphasize that the Brownian motion occurs in the Hilbert space of QM model, hence the gauge potential in the Hilbert space has a meaning of the Berry-type connection and is not related to the any potential in the $x$ space. Recall that under the $T$-duality transform, the "worldvolume" gauge potential along the circle of circumference $R$ gets interchanged with the "target space" coordinate on the dual circle:
\be
A_x\leftrightarrow x, \qquad  R\leftrightarrow \frac{\alpha'}{R}
\ee
where $\alpha'$ is the massive parameter related to the string tension in the string theory framework. Therefore, the electric field gets interchanged with the velocity under the $T$-duality
\be
E=\frac{dA_x}{dt} \quad \leftrightarrow \quad v_x = \frac{dx}{dt}
\ee
since the time is not touched by transform. The object wrapped around the circle under the $T$-duality gets interchanged with the object localized at the dual circle. In particular the Born-Infeld action for the wrapped string gets interchanged with the action of a relativistic particle
\be
\sqrt{1-\alpha' E^2} \quad \leftrightarrow \quad \sqrt{1-v^2}
\ee
How these standard arguments can be applied to our study? Let us assume that the $K \times K$ matrix we are considering, corresponds upon $\frac{\pi}{4}$ rotation to the discretization of the (1+1)D space, which is the target space of Dyck paths in the external magnetic field. Denote the coordinate along the aside diagonal as $x$ and the coordinate along the diagonal as "time", $t$. (equivalently we could interchange them). The coordinate $x$ belongs to the interval $[-K,K]$ and we can assume the periodicity in this coordinate. The constant transverse magnetic field implies the gauge connection along this coordinate in a particular gauge $A_x= Ht$. Another gauge is $A_0=-Hx$ and in this case we treat the aside diagonal as "time coordinate". Upon the $T$-duality transform, the gauge connection of large initial circle gets transformed into the angular coordinate at small dual circle of circumstance of order $K^{-1}$, which is proportional to the "band width" of the transfer matrix. The tridiagonal form of the super tree transfer matrix reflects the smallness of the dual circle. Remind that the size of the matrix, corresponding to the energy level in terms of the $\beta$ ensemble, is $K=\beta N$. We obtain upon the duality
\be
p(k)=ak \quad \leftrightarrow \quad A_x= Ht
\ee
which means that the branching $p$ and the gauge potential are the dual variables under $T$-duality and indeed the branching velocity coincides with the magnetic field. Since the branching $p$ is the angular coordinate on the small dual circle, the value of $k$ in $p(k)=ak$ can be qualitatively
treated as the winding number $\phi =2\pi k$. Note that the interpretation of the branching of the tree as the target space coordinate, or equivalently, as the scalar field, could be useful for the holographic interpretations of super-growing trees.

Can we fit this picture with the ODE/IS duality in the CFT framework \cite{blz,blz2}? It was argued there that the spectral problem in quantum mechanics and the VEV of Baxter operator in the CFT are closely related with the Brownian motion of a (1+1)D particle in the periodic external potential
\be
U(x)= \chi \cos \left(\Phi_{b} + Vx\right)
\ee
where $\Phi_b$ is the boundary value of the (1+1)D scalar field, $\Phi$. The quantum scalar field provides the random environment for the Brownian particle. The argument of the $\cos$-function corresponds to the gauge potential of a constant electric field which has the same meaning for Dyck paths. The parameter $V$ in the potential is related to the parameters in quantum mechanics
as follows \cite{blz}
\be
\left[-\frac{d^2}{d^2x}  + x^{2m} + \frac{l(l+1)}{x^2}\right]\Psi_E(x)= E\Psi_E(x), \qquad
l=\frac{-2iV}{\beta^2} - \frac{1}{2} \qquad m=\frac{1}{\beta^2} -1
\ee
Now we can see which value of the external magnetic field corresponds to the particular potential. To get the oscillator and hence $l=0$, we have to choose
\be
-iV_{osc}= \frac{\beta^2}{4}
\ee
This demonstrates that in our study of Hermite polynomials in the super tree picture, the non-vanishing branching velocity has occurred. The limit of the small external field $V\ll 1$ corresponds to the particular value of  "angular momentum" term in the QM potential . It would be desirable to check that the recursion relations for wave functions corresponding to the QM potential for generic $(m,l)$, can be described by the super trees with the corresponding value of the branching velocity similar to the Hermite polynomials for $(1,0)$ QM.

\subsection{Super trees and torus knot polynomials}

Here we make a remark concerning the interplay between super trees and torus knot invariants. We present the explicit expressions for the parameters of family of QM potentials $(m,l)$ and CFT \cite{blz}
\be
c= 1-\frac{6m^2}{m+1}, \qquad \Delta= \frac{(2l+1)^2 - 4m^2}{16(m+1)}
\ee
where $\Delta$ is the highest-weight of the Virasoro module. The oscillator potential corresponds to the $c=-2$ logarithmic CFT and the free fermion point from the viewpoint of the statistical system.

The link with the torus knot invariants goes through their representation via weighted Dyck paths. Namely, the HOMFLY polynomials $H_{n,n+1}(b,q)$ of the $T_{n,n+1}$ torus knots can be expressed in terms of the weighted Dyck paths in the $n\times n$ square above the diagonal as follows \cite{gor1}
\be
H_{n,n+1}(b,q)= \sum_{Dyck}q^{A}b^{C}
\ee
where $A$ is the area below the path and $C$ is the number of corners on the path. Hence we can link HOMFLY invariants with the super trees via the Dyck paths. In our case the size of the lattice corresponds to the level in the QM spectrum hence the $(n,n+1)$ knot is related to the $n$th energy level. We have no corner counting in our study, hence the relevant object is the lowest row, $b=0$, of the HOMFLY polynomial of the $(n,n+1)$ torus knot in the fundamental representation. It is expressed in terms $q$-deformed Catalan numbers $C_n(q)$ (see, for instance discussion in \cite{bgn})
\be
H_{n,n+1}(b,q)= \sum_{k}b^k\, A_{k}(q), \qquad A_0(q)=C_n(q)
\ee
It is natural to conjecture that the $(n,n+1)$ torus knot invariant evaluates the weighted multiplicity of the corresponded energy level in QM. Indeed there exists example of representation of the HOMFLY invariants of the torus knot as the multiplicity of the $E=0$ states in the Calogero potential although with the different mapping of parameters \cite{gor2} which could be useful along this line of reasoning.

Since our QM has  interplay with CFT it is natural to ask if these CFT data are consistent with the representation of HOMFLY invariant of the knot $K$ in terms of the VEV of Wilson loop in the $SU(N)$ Chern-Simons theory at the level $k$
\be
H_{K}(b,q) = \int D\{A\}\, P\exp\left\{\oint_{K}Adx+ikS_{CS}(A)\right\}
\ee
where $q=e^{\frac{2\pi i}{k+c_v}}$, $b=q^N$, and $P$ is ordering operator. In our study $q\to 1$ limit corresponds to the semiclassical $c\to \infty$ limit in CFT while at the CS representation it corresponds to the semiclassical limit $k\to \infty$ as well.

\section{Random matrices in a Dumitriu-Edelman setting}

I. Dumitriu and A. Edelman have shown in \cite{edelman} that the spectral statistics of various matrix ensembles coincides with the spectral statistics of appropriately chosen ensembles of symmetric tri-diagonal matrices with random independently distributed matrix elements, uniformly distributed along the main diagonal, while non-uniformly distributed along two sub-diagonals. In particular, the spectral density of the Gaussian Orthogonal Ensemble (GOE) coincides with the spectral density of the ensemble of tri-diagonal symmetric matrices of the following form
\be
\hat{M}=\left(\begin{array}{cccccc}
a_{11} & b_{12} & 0 & 0 & 0 & \dots  \medskip \\
b_{21} & a_{22} & b_{23} & 0 & 0 & \medskip  \\
0 & b_{32} & a_{33} & b_{34} & 0 & \medskip \\
0 & 0 & b_{43} & a_{44} & a_{45} & \medskip \\
0 & 0 & 0 & b_{54} & a_{55} &  \medskip \\
\vdots & & & &  & \ddots
\end{array}
\right)
\label{b01}
\ee
where the diagonal elements $a_{kk}$ ($k=1,...,K$) are distributed with the normal distribution, $N(\mu,\sigma)$, while the sub-diagonal elements $b_{k,k+1}\equiv b_{k+k,i}$  ($k=1,...,K$) share the $\chi_{k}$-distribution. The normal and the $\chi$-distributions have the following probability densities for a random value, $x$, representing the matrix element:
\be
\left\{\begin{array}{rcll}
f(x|\mu,\sigma) & = & \disp \frac{1}{\sqrt{2\pi}\sigma}e^{-\frac{(x-\mu)^2}{2\sigma^2}} & \quad \mbox{for a normal distribution} \medskip \\
f(x|n) & = & \disp \frac{x^{n-1}e^{-\frac{x^2}{2}}}{2^{\frac{n}{2}-1} \Gamma\left(\frac{n}{2}\right)},\quad x\geq 0 & \quad \mbox{for a $\chi$-distribution}
\end{array}\right.
\label{b01a}
\ee
where $\Gamma(z)$ is the Gamma-function.

The symmetric matrix \eq{b01} (where $b_{ij}=b_{ji}$) allows a straightforward interpretation as the transfer matrix of a path counting problem on a random symmetric super tree. To proceed, introduce the "shifted" matrix $\hat{M}'$, where
\be
\hat{M}'=\left(\begin{array}{cccccc}
a_{11} & 1 & 0 & 0 & 0 & \dots  \medskip \\
b^2_{21} & a_{22} & 1 & 0 & 0 & \medskip  \\
0 & b^2_{32} & a_{33} & 1 & 0 & \medskip \\
0 & 0 & b^2_{43} & a_{44} & 1 & \medskip \\
0 & 0 & 0 & b^2_{54} & a_{55} & \medskip \\
\vdots & & & & & \ddots
\end{array}
\right)
\label{b01c}
\ee
We can immediately see that for any distribution of matrix elements
$$
\det \hat{M} = \det \hat{M}'
$$
Thus, we can deal with the matrix $\hat{M}'$ and consider it as a transfer matrix of a \emph{random super tree} constructed as follows:
\begin{itemize}
\item[(i)] All nodes at the generation $k$ of a super tree carry one and the same random $N(\mu,\sigma)$-distributed weight;
\item[(ii)] The branching (vertex degree) of all vertices at the generation $k$ of a super tree  is a $\chi_k$-distributed random variable.
\end{itemize}
For the path counting problem on thus defined random super tree, the condition (i) gives the normally distributed diagonal matrix elements, while the condition (ii) ensures the correct $\chi$-distributed weights for passages between adjacent generations of the tree (from $k$ to $k+ 1$).

Now, let us find a "mean tree", where the averaging is taken over the ensemble of random trees. It can be easily seen that such a "mean tree" is nothing else as the super-growing tree discussed at length of our paper. Instead of considering the spectral density of the ensemble of random matrices $\hat{M}$, we pay attention to the eigenvalue distribution of the mean matrix $\la \hat{M} \ra$, obtained by replacing each matrix element of $\hat{M}$ by its mean value. The mean values of all diagonal elements are $0$ since the probability density, $f(x|\mu,\sigma)$, is symmetric at $\mu=0$, while the mean values (the expectations) of off-diagonal random elements are given by the following expression
\be
\mathbf{E}_{\chi_{(k)}}(x)= \frac{\sqrt{2}\,\Gamma\left(\frac{k+1}{2}\right)}{\Gamma\left(\frac{k}{2}\right)}
\label{b02}
\ee
For $k\gg 1$ the expectation $\mathbf{E}_{\chi_{(k)}}(x)$ has the asymptotic expression
\be
\mathbf{E}_{\chi_{(k)}}(x)\big|_{k\gg 1} = \sqrt{k}
\label{b03}
\ee
Thus, the averaged matrices $\la \hat{M}\ra$ and $\la \hat{M}'\ra$ have the following forms
\be
\la \hat{M}\ra\approx \left(\begin{array}{cccccc}
0 & \sqrt{1} & 0 & 0 & 0 & \dots  \medskip \\
\sqrt{1} & 0 & \sqrt{2} & 0 & 0 & \medskip  \\
0 & \sqrt{2} & 0 & \sqrt{3} & 0 & \medskip \\
0 & 0 & \sqrt{3} & 0 & \sqrt{4} & \medskip \\
0 & 0 & 0 & \sqrt{4} & 0 & \medskip \\
\vdots & & & & & \ddots
\end{array}
\right); \quad
\la \hat{M}'\ra\approx \left(\begin{array}{cccccc}
0\;\; & 1\;\; & 0\;\; & 0\;\; & 0\;\; & \dots  \medskip \\
1\;\; & 0\;\; & 1\;\; & 0\;\; & 0\;\; & \medskip  \\
0\;\; & 2\;\; & 0\;\; & 1\;\; & 0\;\; & \medskip \\
0\;\; & 0\;\; & 3\;\; & 0\;\; & 1\;\; & \medskip \\
0\;\; & 0\;\; & 0\;\; & 4\;\; & 0\;\; & \medskip \\
\vdots & & & & & \ddots
\end{array}
\right)
\label{b04}
\ee
Since $\det \la \hat{M} \ra = \det \la \hat{M}' \ra = \det \hat{T}$, where $\hat{T}$ is the transfer matrix of the super tree defined in \eq{03}, we can interpret the spectral statistics of path counting on super trees as a suitable mean-field model describing spectral properties of random matrix ensembles, which captures the KPZ scaling near the spectral edge.

\section{Conclusion}

We have touched at length of this paper some aspects of a path counting  problem on "super trees" ${\cal T}^{\pm}$ whose branching linearly grows or decreases with the generation of the tree. It was argued that such super trees emerge naturally in some statistical models or CFT describing the Hilbert space of some QM problems. This has been demonstrated in the simplest oscillator case, in which the super-growing tree with the "velocity" $a=1$ emerges naturally. It would be interesting to get the super-growing tree determinant representation for another potentials. Generically the super-growing trees which reflect the recurrence between the wave functions involve more complicated "time dependent" branching velocities. The super tree with small branching velocity turns out to be related via a kind of $T$-duality transform with the area-weighted Dyck paths and therefore with some particular case of HOMFLY polynomials for torus knots.

It is worth mentioning few immediate questions for further research. First, note that a new type of  spin chains has been formulated recently in \cite{korepin1,korepin2,fradkin} in which the interaction term in the Hamiltonian involves spins at three neighboring sites. The ground states of such spin chain \cite{korepin1} are in one-to-one correspondence with Dyck paths, while
the ground states in the generalized spin chain model \cite{korepin2,fradkin} are in bijection with the area-weighted Dyck paths. It would be interesting to recognize the place of the super-growing tree in the generalized spin chain model and to describe their entanglement properties in terms of statistics of paths on super trees.

Another interesting issue concerns the possible holographic role of super-growing trees. The standard Cayley tree is used for the modeling the hyperbolic 2D geometry. The super-growing tree certainly modifies $AdS_2$ geometry. Since the parameter of modification corresponds to the coefficient in front of the $r^{-2}$--term in the QM potential, it is worth reminding that such a potential emerges for a particle nearby the black hole horizon. Hence one could speculate that the super-growing tree could be relevant for the discretization of the metric of BH in $AdS_2$. Such a metric in Jackiw-Teitelboim gravity implies that the branching number should be related to the value of the effective 2D dilaton field with linear behavior in radial coordinate.

The key property used in our study is the possibility to develop the simple path representation of the Hilbert space of the one-body QM models. Certainly this property is due to the representation of the Hilbert space in terms of the group representation. This is familiar property of all QM models with algebraization of the spectrum. The same analysis and the path representation of the wave functions of the integrable many-body systems suggests that the degeneration of the spectrum can be related with the knot invariants via the super-tree -- Dyck path correspondence in a more general situation.

It would be interesting to develop representation for all QM ingredients is terms of the
trees with varying degrees. In particular it is interesting to get the tree representation
for the Wigner function defined on the phase space, Moyal product on the trees,
Witten index in SUSY QM, matrix
elements of the different operators etc. We shall discuss these issues elsewhere.

Another issue providing some geometrical interpretation of path counting on super trees deals with a deep connection of this model with the spectral determinant of Gaussian Orthogonal Ensembles. This connection sheds some light on physically puzzling interpretation of integrals over random GOE matrices as expectations of special random tri-diagonal matrices taken from non-uniform $\chi$-ensemble, discovered by A. Edelman and I. Dumirtiu \cite{edelman}.

One more possible development concerns the interpretation of the 1D Kardar-Parisi-Zhang scaling with the critical exponent $\nu=\frac{1}{3}$. Let us look at KPZ-type scaling which appears in our model from a slightly different perspective. We might be interested in manifestations of a KPZ-type scaling in other physical phenomena, happen (at least) in physics of one-dimensional disordered systems. In particular, we could ask whether there is a connection of the one-dimensional Anderson localization with the KPZ-type scaling behavior? It seems that the answer is positive. To provide the idea behind this connection, let us recall that the behavior of the density of states, $r(E)$, of the one-dimensional Anderson model (tight-binding model with randomness on the main diagonal) at $E\to 0$, has the asymptotics known as the "Lifshitz singularity"
\be
r(E) \sim e^{-\frac{\alpha}{\sqrt{E}}}
\ee
where $E$ is the energy of the system and $\alpha$ is some positive constant (see \cite{lif1,lif2} for more details). Consider the canonical ensemble, in which $E$ is controlled only in average by the conjugated Legendre variable, $N$. In this case the density of states, $r(E)$, gets converted into $r(N)$ via the Laplace transform:
\be
r(N) = \left.\int_0^{\infty} r(E)\, e^{-N E}\, dE\;\right|_{L\gg 1} \sim \varphi(N)\, e^{-\left(\frac{3\alpha}{2}\right)^{2/3}\,N^{1/3}}
\label{lap}
\ee
where at $N\gg 1$ we pay attention to the exponential asymptotics only and neglect the power law corrections such as $\varphi(N)\sim N^{-5/6}$. Correspondingly, the density of states, $r(E)$, can be restored from $r(N)$ via the inverse Laplace transform. Let us associate $r(N)$ with the partition function $Z_N^{(W)} = e^{S_N^{(W)}}$ defined in \eq{09b}--\eq{09c} (which is meaningful since the density of states plays the role of a properly normalized partition function). Substituting $Z_N^{(W)}$ for $r(N)$, we get
\be
r(E) = \frac{1}{2\pi i} \int_{\gamma-i\infty}^{\gamma+i\infty} e^{S_N^{(W)}}\, e^{N E}\, dN  \approx \frac{1}{2\pi i} \int_{\gamma-i\infty}^{\gamma+i\infty} e^{a_1 N^{1/3}}\, e^{N(E+c)}\, dN \sim e^{- \frac{\beta}{\sqrt{E+c}}}
\ee
where $\beta = \frac{2}{3} |a_1|^{3/2}$ (recall that $a_1\approx-2.33811$) and the presence of of the linear term $cN$ in the entropy $S_N^{(W)}$ shifts the singularity of the function $r(E)$ to the value $-c$.

\begin{figure}[ht]
\epsfig{file=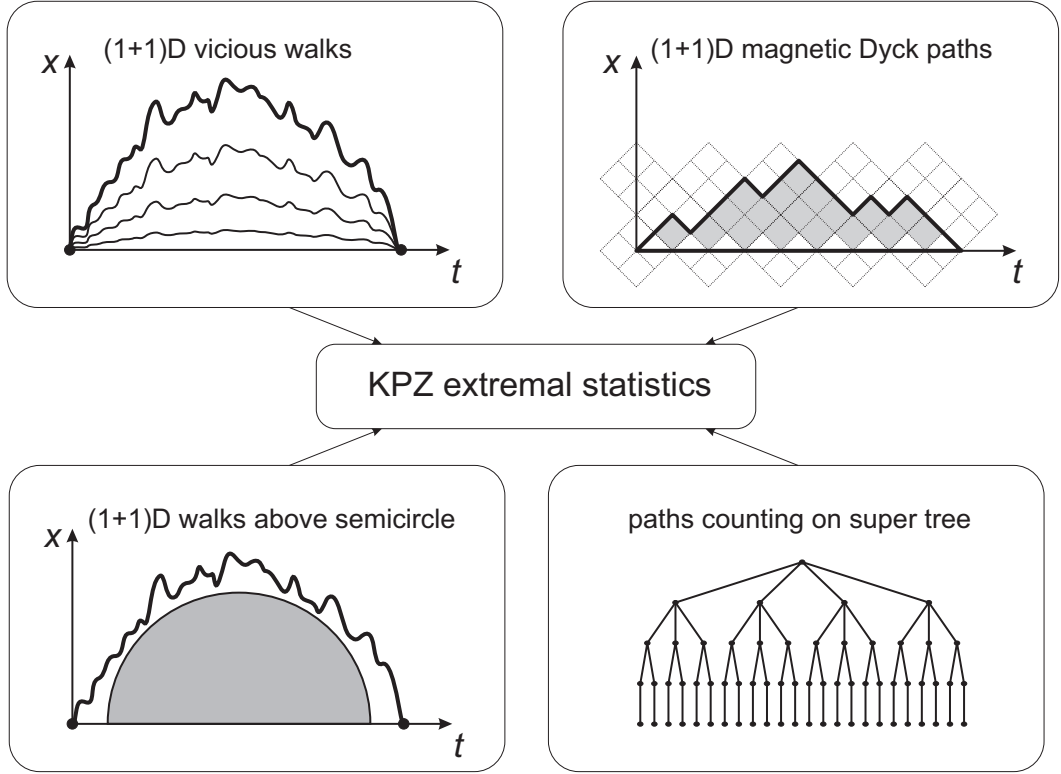, width=14cm}
\caption{KPZ scaling, appearing in a set of interconnected models: vicious walks, directed paths evading the semicircle, "inflated" magnetic Dyck paths, closed paths on descending super tree.}
\label{fig:05}
\end{figure}

It is known \cite{neuwen} that the leading exponential asymptotic \eq{lap}, appeared in the literature under various names, like "stretched exponent", "Griffiths singularity", "Balagurov-Waks trapping exponent", is nothing else as the Laplace-transformed Lifshitz tail of one-dimemsional disordered systems possessing Anderson localization. We claim that KPZ-type behavior can also be regarded as an incarnation of a specific "optimal fluctuation" for one-dimensional Anderson localization. The appearance of KPZ scaling in a set interconnected models is schematically shown in \fig{fig:05}. The fluctuations of the top line in a bunch of vicious walks in the mean-filed setting can be modelled either by the random walk statistics above the impenetrable semicircle, or by sufficiently "inflated" magnetic Dyck paths, or by statistics of closed paths on descending super tree. In all mentioned cases the KPZ statistics occurs when the system is pushed to the "large deviation" region, i.e. the KPZ behavior is seen in the very untypical sub-ensemble of the ensemble of whole available paths. In pore details some examples will be discussed at length of the forthcoming paper \cite{nech_pol_val}.

One more remark concerning the possible relevance of our study for the issue of Anderson localization goes as follows. It was suggested in \cite{kamenev} that the problem of Anderson localization for the many-body systems with interaction can be translated into the one-particle Anderson localization in the Hilbert space of the interacting system. The idea was to approximate the Hilbert space of the many-body system by the Bethe tree and use the exact results concerning
the one-particle localization on the tree \cite{anderson}. It was shown later \cite{basko} shown that the many-body localization which prevents the thermalization of the interacting systems can be formulated within this approach.

In this note we have built the tree which captures the information about the recursion properties of the wave functions for the $(m,l)$ family of the QM models. These models can be considered as a particle in the external potential or the relative motion of two interacting particles. Our super trees provide the proper parametrization of the Hilbert space of the QM system in sense of \cite{kamenev}. However, it turns out that the tree is not a Bethe tree even for the simplest Hilbert space of the harmonic oscillator- it is the tree with the varying branching of  nodes. In the $(m,l)$ series of potentials besides the $l=-\frac{1}{2}$ case, the super trees govern the Hilbert spaces and the branching velocity is fixed by the coefficient in front of the $x^{-2}$ term in the potential. Curiously the branching of the tree itself behaves like a scalar field with
particular radial behavior.

Hence, in our sample model the localization of the single degree of freedom in the interacting system is more complicated because the tree generically is not homogeneous. The randomness in the initial potential generically could yield apart from the simple on-site disorder the randomization of the valencies of the nodes. Our example of the constant branching velocity seems to be one of the simplest cases. Remark that the wave function of the QM mechanics corresponds to the characteristic polynomial of the transfer matrix of statistical model on the tree hence the localization properties of the particle in the physical space encoded in the inverse participation ration gets mapped precisely into the properties of the moments of the spectral density of the matrices of the walking at the super tree.

\begin{acknowledgments}
We are grateful to V. Avetisov, M. Tamm and A. Kamenev for the useful discussions and remarks.
The work of A.G. was performed at the Institute for Information Transmission Problems with the financial support of the Russian Science Foundation (Grant No.14-50-00150); S.N. acknowledges the support of the EU-Horizon 2020 IRSES project DIONICOS (612707), and of the RFBR grant No. 16-02-00252.
\end{acknowledgments}


\begin{thebibliography}{99}

\bibitem{tateo} P. Dorey and R. Tateo, J. Phys. A {\bf 32} L419 (1999)

\bibitem{blz} V.V. Bazhanov, S.L. Lukyanov, and A.B. Zamolodchikov, J. Stat. Phys. {\bf 102} 567 (2001)

\bibitem{suzuki} J. Suzuki, J. Phys. A 32 (1999) L183

\bibitem{rev} P. Dorey, C. Dunning and R. Tateo,  J. Phys. A {\bf 40} R205 (2007)

\bibitem{blz2} V.V. Bazhanov, S.L. Lukyanov, and A.B. Zamolodchikov, Nucl. Phys. B {\bf 549} 529 (1999)

\bibitem{vafa} M. Aganagic, M.C.N. Cheng, R. Dijkgraaf, D. Krefl, and C. Vafa, JHEP {\bf 1211} 019 (2012)

\bibitem{krefl1} D. Krefl, JHEP {\bf 1412} 118 (2014)

\bibitem{krefl2} D. Krefl, JHEP {\bf 1608} 020 (2016)

\bibitem{aga} M. Aganagic, R. Dijkgraaf, A. Klemm, M. Marino, and C. Vafa, Comm. Math. Phys. {\bf 261} 451 (2006)

\bibitem{kpz} M. Kardar, G. Parisi, and Y.-C. Zhang, Phys. Rev. Lett. {\bf 56} 889 (1986)

\bibitem{halpin} T. Halpin-Healy and Y.-C. Zhang, Phys. Rep. {\bf 254} 215 (1995)

\bibitem{johansson} K. Johansson, Comm. Math. Phys. {\bf 209} 437 (2000)

\bibitem{spohn} M. Pr\"ahofer and H. Spohn, Phys. Rev. Lett. {\bf 84} 4882 (2000); M. Pr\"ahofer and H. Spohn, J. Stat. Phys. {\bf 108} 1071 (2002)

\bibitem{tw} C.A. Tracy and H. Widom, Commun. Math. Phys. {\bf 159} 151 (1994)

\bibitem{bgn} K. Bulycheva, A. Gorsky and S. Nechaev, Phys. Rev. D {\bf 92}, 105006 (2015)

\bibitem{17} A. Maritan, Phys. Rev. Lett. 62 2845 (1989)

\bibitem{ternovsky} F.F. Ternovsky, I.A. Nyrkova, and A.R. Khokhlov, Physica A {\bf 184} 342 (1992)

\bibitem{burda} Z. Burda, J. Duda, J.-M. Luck, and B. Waclaw, Phys. Rev. Lett. {\bf 102} 160602 (2009)

\bibitem{heavy} S.K. Nechaev, M.V. Tamm, and O.V. Valba, J. Stat. Mech. 053301 (2017)

\bibitem{Kornyik} M. Kornyik and G. Michaletzky, J. Approx. Theor. {\bf 211} 29 (2016)

\bibitem{Dominici} D. Dominici, J. Diff. Eq. Appl. {\bf 13} 1115 (2007)

\bibitem{edelman} I. Dumitriu and A. Edelman, J. Math. Phys. {\bf 43} 5830 (2002)

\bibitem{carlitz} L. Carlitz and J. Riordan, Duke J. Math. {\bf 31} 371 (1964); J. F\"urlinger and J. Hofbauer, J. Comb. Theor. A {\bf 40} 248 (1985)

\bibitem{prel0} T. Prellberg and R. Brak, J. Stat. Phys. {\bf 78} 701 (1995)

\bibitem{rich1} C. Richard and A. J. Guttmann, and I. Jensen, J.Phys. A: Math. Gen. {\bf 34} L495 (2001)

\bibitem{rich2} C. Richard, J. Stat. Phys., {\bf 108} 459 (2002)

\bibitem{lif1} I. M. Lifshitz, Sov. Phys. JETP, {\bf 26} 462 (1968)

\bibitem{lif2} I. M. Lifshitz, S. A. Gredeskul, and L. A. Pastur, \emph{Introduction to the theory of disordered systems} (Wiley-Interscience: 1988)

\bibitem{neuwen} Th. M. Nieuwenhuizen, Phys. Rev. Lett. {\bf 62} 357 (1989)

\bibitem{gor1} E. Gorsky, Zeta functions in algebra and geometry (2012), 213-232, e-Print: arXiv:1003.0916

\bibitem{gor2} E. Gorsky, Selecta Mathematica, New Series (2013), 1-16, e-Print: arXiv:1110.3524

\bibitem{nech_pol_val} S. Nechaev, K. Polovnikov, A. Valov, to be published

\bibitem{kamenev} B. Altshuler, Y.Gefen, A. Kamenev, and L. Levitov, Phys. Rev. Lett. {\bf 78} 2803 (1997)

\bibitem{anderson} R. Abou-Chacra, D.J. Thouless, and P.W. Anderson, J. Phys. C {\bf 6}
1734 (1973)

\bibitem{basko} D. Basko, I. Aleiner, and B. Altshuler, Annals of Physics {\bf 321} 1126 (2006)

\bibitem{korepin1} O. Salberger and V. Korepin, Rev. Math. Phys. {\bf 29} (2017) 1750031

\bibitem{korepin2} O. Salberger, T. Udagawa, Z. Zhang, H. Katsura, I. Klich, and V. Korepin, J. Stat. Mech. {\bf 1706}, 063103 (2017)

\bibitem{fradkin} X. Chen, E. Fradkin, and W. Witczak-Krempa, J. Phys. A {\bf 50} 464002 (2017)
	
\end{thebibliography}
\end{document}